\def\b{\beta}

\def\d{\delta}

\def\IR{\relax{\rm I\kern-.18em R}}
\font\cmss=cmss10 \font\cmsss=cmss10 at 7pt
\def\IZ{\relax\ifmmode\mathchoice
{\hbox{\cmss Z\kern-.4em Z}}{\hbox{\cmss Z\kern-.4em Z}}
{\lower.9pt\hbox{\cmsss Z\kern-.4em Z}}
{\lower1.2pt\hbox{\cmsss Z\kern-.4em Z}}\else{\cmss Z\kern-.4em Z}\fi}
\def\IN{\relax{\rm I\kern-.18em N}}

\documentstyle[preprint,aps]{revtex}

\newcommand{\bge}{\begin{equation}}
\newcommand{\ege}{\end{equation}}
\newcommand{\bga}{\begin{eqnarray}} 
\newcommand{\ega}{\end{eqnarray}}

\begin{document}
\draft
\title{Effect of electron correlation on superconducting pairing symmetry}
\author{Manidipa Mitra}
\address{Institute of Physics, Sachivalaya Marg, Bhubaneswar - 751 005, India.}
\author{Haranath Ghosh}
\address{Instituto de Fisica, Universidade Federal Fluminence, Campus
da praia Vermelha, Av. Litor$\hat a$nea s/n, 24210 - 340, Niteroi RJ, Brazil.}
\author{S. N. Behera}
\address{Institute of Physics, Sachivalaya Marg, Bhubaneswar - 751 005, India.}
\maketitle
\begin{abstract}
The role of electron correlation on 
different pairing symmetries are discussed in details where the electron
correlation has been treated within the slave boson formalism. 
It is shown that for a pure $s$ or pure $d$ wave pairing symmetry,
the electronic correlation suppresses the $s$ wave gap magnitude 
(as well as the $T_c$) at a faster rate than that for the $d$ wave
gap. On the otherhand, a complex order parameter of the form ($s+id$) shows 
anomalous temperature dependence. For example, if the temperature ($T_{c}^d$) at
which the $d$ wave component of the complex 
order parameter vanishes happens to be 
larger than that for the $s$ wave component ($T_{c}^s$) then the growth of the
$d$ wave component is arrested with the onset of the $s$ wave component of
the order parameter. In this mixed phase however, we find that 
the suppression in different components of the gap as well as the 
corresponding $T_c$ due to coulomb correlation
 are very sensitive to the relative pairing strengths
of $s$ and $d$ channels as well as the underlying lattice.
Interestingly enough, in such a scenario (for a case of $T_{c}^s > T_{c}^d$)
the gap magnitude of the $d$ wave component increases with electron 
correlation but not $T_{c}^d$ for certain values of electron correlation. 
However, this never happens in case of the 
$s$ wave component.
We also calculate the temperature dependence of the 
superconducting gap along both the high symmetry directions ($\Gamma$ - M and
$\Gamma $ - X) in a mixed $s+id$ symmetry pairing state  
and the thermal variation
of the gap anisotropy ($\frac{\Delta_{\Gamma - M}}{\Delta_{\Gamma - X}}$) 
with electron correlation. The results are discussed with reference to
experimental observations.
\end{abstract}
{\bf PACS. No. : } 74.25.DW, 74.62.-c, 74.20.Fg. 
\section{Introduction}
The correlated electron
systems like the high temperature cuprate superconductors, 
show various anomalous physical properties in the normal as well
as in the superconducting state. Alongwith the unsolved problems 
concerning the pairing
mechanism in these exotic materials the question of order parameter symmetry is 
also not yet understood. In the weak coupling conventional BCS superconductors,
superconductivity results from a pairing of electrons via  phonon-
mediated  attractive 
electron-electron interaction which dominates the usual Coulomb
repulsion at low temperature, and an energy gap $\Delta (k)$ appears in the 
quasiparticle spectrum. This energy gap is very nearly isotropic in $k$-space,
so that the gap has the same magnitude and phase in all directions i.e., 
an isotropic s-wave pairing state. In the high-$T_c$ superconductors 
the pairing of charge carriers is established by experiments on flux 
quantization \cite{4.1}, I-V characteristics
of Josephson tunnel junction \cite{4.2} and Andreev scattering \cite{4.3}.
But the point contact tunneling \cite{4.4} on various high-$T_c$ materials 
and nuclear quadrupole/magnetic resonance and relaxation measurements \cite{4.5}
conclude that the energy gap function in these materials 
has considerable anisotropy. In high-$T_c$ superconductors, 
measurements of the temperature dependence of the NMR Knight shift 
\cite{4.5,4.6} of suitable 
nuclei in a superconductor ruled out the possibility of p-wave pairing but whether the pairing
is of s-wave or d-wave type is not yet settled. The results of NMR experiments
measuring Cu-relaxation rates by Martindale et al \cite{4.7}, are in agreement
with the prediction of d-wave pairing. Measurement of penetration depth $\lambda$ at low temperatures \cite{4.8} on $YBa_2Cu_3O_{6.95}$ supports d-wave pairing.
Recent angle - resolved photoemission spectroscopy (ARPES) study \cite{4.8} also suggests
a d-wave state. However, there exist some experimental results like variation
of penetration depth in $Nd_{2-x}Ce_xCuO_4$ \cite{4.9}, measurements of 
Josephson supercurrent for tunneling between Pb and $YBa_2Cu_3O_7$ \cite{4.10}
which does not correspond to the
 exact d-wave symmetry. Moreover, some photoemission studies
on $Bi_2Sr_2CaCu_2O_{8+x}$ \cite{4.11} are inconsistent with pure d-wave but are
more consistent with a mixed state of s and d wave components. Kotliar \cite{4.12}
and Ruckenstein et al, \cite{4.13} first introduced the concept of mixed $s+id$
and $s+d$ symmetries respectively. Their ideas of mixed configuration was 
to interpret
the NMR and NQR data in the superconducting state of $YBCO$ \cite{4.14} and
the Josephson critical current measurements in $YBCO$ (superconductor-normal-superconductor) SNS junctions 
and YBCO/Pb junction \cite{4.15}. 
\par Based on the discussions on the experimental situation about the 
determination of the order parameter symmetry in high temperature 
superconductors it is clear that there are various opinions about 
pairing symmetry in cuprates, which are known to be correlated system.
Therefore, in the present paper the effect of Coulomb correlation on the 
superconducting state is considered based on 
a weak coupling theory. The role of electron correlation on 
different pairing symmetries are discussed in details where the electron
correlation has been treated within the slave boson formalism. 
It is shown that for a pure $s$ or pure $d$ wave pairing symmetry,
the electronic correlation suppresses the $s$ wave gap magnitude 
(as well as the $T_c$) at a faster rate than that for the $d$ wave
gap. On the otherhand, a complex order parameter of the form ($s+id$) shows 
anomalous temperature dependence. For example, if the temperature ($T_{c}^d$) 
at which the $d$ wave component of the order parameter vanishes happens to be 
larger than that for the $s$ wave component ($T_{c}^s$) then the growth of the
$d$ wave component is arrested with the onset of the $s$ wave component of
the order parameter. In this mixed phase 
 however, it is shown that 
the suppression in different components of the gap as well as the 
corresponding $T_c$ are very sensitive to the relative pairing strengths
of $s$ and $d$ channels as well as the underlying lattice.
Interestingly enough, in such a scenario (for a case of $T_{c}^s > T_{c}^d$)
the zero temperature gap magnitude of the $d$ wave component increases with electron 
correlation but not $T_{c}^d$ for certain values of electron correlation. 
However, this never happens in case of the 
$s$ wave component. 
We also calculate the temperature dependence of the 
superconducting gap along both the high symmetry directions ($\Gamma$ - M and
$\Gamma $ - X) in a mixed $s+id$ symmetry pairing state  and 
the thermal variation
of the gap anisotropy ($\frac{\Delta_{\Gamma - M}}{\Delta_{\Gamma - X}}$) 
with electron correlation. The results are discussed with reference to
experimental observations.

\par The lay out of the rest of the paper is as follows. In section II we describe the effect 
of on-site coulomb correlation on weak coupling superconductivity. In this section we also present the slave boson
approximation to the on-site coulomb correlation in the Hubbard model. The next section III is devoted to describe
the effect of electronic correlation on different order parameter symmetries. Effect of electron correlation on each 
order parameter symmetry is described sequentially in each subsection of this section with detailed discussions on 
the results. A comparison of the present study with ARPES results in high-$T_c$ materials are described in section IV.
Finally, remarks have been made about the present calculation in the conclusion section V. 
\section {Effect of electron correlation on a weak coupling theory of superconductivity}

 The simplest model that describes the electron-electron interaction 
in a correlated
system, like the high-$T_c$ cuprates, is the single band Hubbard model.
One of the major problems in solving 
the Hubbard model is how to treat the correlation exactly. 
One of the ways to do
so is to take recourse to the slave boson formulation 
of Kotliar and Ruckenstein (KR) \cite{kot}. 
In this representation a set of four bosons are assigned for four possible 
occupancies of a lattice site. 
The bosonic fields which keep track of the
four different occupations of a site `$i$' are $e_{i}^{\dagger}(e_i)$,
$ s_{{i\sigma}/(-\sigma)}^\dagger ( s_{{i\sigma}/(-\sigma)})$ and 
$d_i^\dagger(d_i)$ corresponding to
creation (annihilation) of empty, single occupation with 
spin $\sigma /(-\sigma)$ and double occupation
respectively. Of course the total probability 
of occupation of any site is one
and it has to be respected. Secondly, the fermion number should be conserved at
`$i$'th site in this
slave boson representation. These two constraints are imposed by two conditions,
 completeness relation
and charge conservation, written as,
\begin{equation}
e_i^\dagger e_i + \sum_{\sigma}s_{i\sigma}^\dagger s_{i\sigma} + d_i^\dagger d_i
 = 1
\label{2.22}
\end{equation}
\begin{equation}
c_{i\sigma}^\dagger c_{i\sigma} = s_{i\sigma}^\dagger s_{i\sigma} + d_i^{\dagger
}d_i
\label{2.23}
\end{equation}
The Hubbard Hamiltonian is described as
\begin{equation}
H = \sum_{ij\sigma} t_{ij} c_{i\sigma}^\dagger c_{j\sigma} + U\sum_i 
n_{i \uparrow}n_{i\downarrow}
- \mu \sum_{i\sigma} n_{i\sigma}
\label{2.24}
\end{equation}
where the first term is the usual tight binding Hamiltonian with 
the hopping integral $t_{ij}$, the
second term denotes the intra-atomic Coulomb repulsion ($U$) between 
two electrons with opposite spins
occupying the same site and the third term denotes the chemical 
potential ($\mu $) which can take into
account the deviation from half - filling. The prescription for 
the transformation of
equation \ (\ref{2.24}) in terms of
the slave boson operators is given below :
\bga
\left .
\begin{array}{ll}
n_{i\uparrow}n_{i\downarrow} \longrightarrow d_i^\dagger d_i
\\
n_{i\sigma} \longrightarrow \tilde n_{i\sigma}
\\
c_{i\sigma}^\dagger c_{j\sigma} \longrightarrow z_{i\sigma}^\dagger \tilde
c_{i\sigma}^\dagger \tilde c_{j\sigma} z_{j\sigma}
\end{array}
\right \}
\label{2.25}
\ega
where
\begin{equation}
z_{i\sigma} = (1- d_i^{\dagger} d_i - s_{i\sigma}^\dagger s_{i\sigma})^{-\frac
{1}{2}} (e_i^\dagger s_{i\sigma} + s_{i-\sigma}^{\dagger} d_i)
(1 - e_i^{\dagger}e_i - s_{i-\sigma}^{\dagger} s_{i-\sigma})^{-\frac{1}{2}}
\label{2.26}
\end{equation}
and $\tilde c_{i\sigma}^\dagger$ is the modified fermion creation operator.
The form of $z_{i\sigma}$ is so chosen as to reproduce the correct band structure
in the absence of correlation. The operators $(e_i^\dagger s_{i\sigma} +
s_{i-\sigma}^\dagger d_i)$ in the  eqn. \ (\ref{2.26})
describe the hopping process of electron i.e., if an electron hops from
site `$i$' to `$j$', the slave bosons must simultaneously change at $j$ and
$i$. Depending on whether the 
site `$i$' is singly or doubly occupied the bosonic
state of `$i$' must change from $s_{i\sigma}^\dagger$ 
to $e_i^\dagger$ or from $d_i^\dagger$
to $s_{i-\sigma }^\dagger $. Thus there are two transition channels which
add up, and the total transition probability must be equal to one. It is
therefore useful to introduce a normalization factor, which guarantees the
conservation of probability even in the mean-field theory.

\par 
 Following KR the Hubbard Hamiltonian 
can be re-written
in terms of the slave boson operators as
\bga
H & = & \sum_{ij\sigma}t_{ij}z_{i\sigma}^{\dagger} \tilde c_{i\sigma}^\dagger
\tilde c_{j\sigma} z_{j\sigma} + U\sum_i d_i^\dagger d_i - \mu \sum_{i\sigma}
\tilde c_{i\sigma}^\dagger \tilde c_{i\sigma}
\nonumber \\
& &
+\sum_{i\sigma} \lambda_{i\sigma} (\tilde c_{i\sigma}^\dagger \tilde c_{i\sigma}
 -
s_{i\sigma}^\dagger s_{i\sigma} - d_{i}^\dagger d_i)
\nonumber \\
& &
+\sum_i \lambda_i^\prime (1 - e_{i}^\dagger e_i - d_i^\dagger d_i - \sum_{\sigma
}
s_{i\sigma}^\dagger s_{i\sigma})
\label{2.27}
\ega
where $\lambda_{i\sigma}$ and $\lambda_i^\prime$ are Lagrange multipliers 
which enforce the charge conservation and completeness relation (1,2)
respectively. 
So the Coulomb interaction term is simplified and diagonalized with double 
occupancy operators but the kinetic energy part becomes complicated.
The values of the boson field operators and the Lagrange multipliers
are determined by minimizing the free energy of the system in the saddle
point approximation, where all the bose operators and Lagrange multipliers
are replaced by c-numbers.
In this approximation the Hamiltonian is
\bga
H & = & \sum_{ij\sigma}\tilde q t_{ij} \tilde c_{i\sigma}^\dagger 
\tilde c_{j\sigma}
+ (\lambda - \mu) \sum_{i\sigma} \tilde c_{i\sigma}^\dagger \tilde c_{i\sigma}
\nonumber \\
& &
+N \left [ Ud^2 - 2\lambda(d^2+s^2) + \lambda^\prime (1 - d^2 - e^2 -2s^2) 
\right ]
\label{2.28}
\ega
where $\tilde q = <z_{i\sigma}^\dagger z_{j\sigma}>$, $e$, $d$, $s$, 
$\lambda $,
$\lambda^\prime$ are the saddle point values of the respective field operators,
`$N$' is the number of sites, and in the case of paramagnetic ground state
$s_{\sigma}=s_{-\sigma}=s$. The Hamiltonian of the system takes the simple
form of an effective tight binding model with a modified hopping integral
of the form $t_{ij} \longrightarrow \tilde q t_{ij}$, where the correlation
effects are built in through the multiplicative factor
$\tilde q = <z^\dagger z>$. $\tilde q$ in general is a complicated function
of the coulomb correlation $u(=\frac{U}{U_c}, ~U_c$ being the Brinkmann-Rice
value for metal-insulator transition) and the dopant concentration $\delta$.
In this approach solutions are obtained for the paramagnetic states for all
values of $u$ and band fillings that reproduces the correct Brinkmann-Rice
result for metal-insulator transition at a critical value of correlation
($U_c$) at half-filling. Approximately, in the weak correlation limit $\tilde
q = 1 -u^2$ and in case of strong correlation and small values of $\d$, 
$\tilde q =\frac{2 \d}{\sqrt{1-u^{-1}}}$ (for details 
please see ref. \cite{a,b}).
\par Now in order to study the effect of electron correlation 
on SC pairing symmetry we use a model
Hamiltonian which contains in addition to a repulsive on-site 
coulomb correlation (7) 
term, a pairing term that leads to superconductivity (for our purpose). 
The pairing 
interaction is assumed to be due to some boson exchange (unknown) mechanism, as
there exists no conclusive pairing mechanism for the 
high-$T_c$ superconductivity. 
The  pairing hamiltonian in general may
be written as,
\bge
H_p=-\sum_{kk^\prime}V_{kk^\prime}
c_{k\uparrow}^\dagger c_{-k\downarrow}^\dagger c_{-k^\prime \downarrow}
c_{k^\prime \uparrow}
\ege
While treating the electron correlations using salve boson formalism,
the nature of the quasiparticles are  no longer 
the bare electrons  but the electronic
quasiparticles of the correlated system (i.e $\tilde c$).
Therefore the bare electronic operators should be transformed accordingly as 
equation (4).
As a consequence, the strength of the pairing interaction
becomes $V_{kk^\prime} \longrightarrow \tilde q^2 V_{kk^\prime}$ when the
$c_{k\sigma}(c_{k\sigma}^\dagger$) are replaced by new quasiparticle operators
$\tilde c_{k\sigma}(\tilde c_{k\sigma}^\dagger$). Hence, the total hamiltonian
for the superconducting state in a correlated system can be written as,
\bge
H=H_0 + H_B + H_{p}
\ege
where
\bge
H_0=\sum_{k\sigma}(\tilde q \epsilon_k - {\mu})\tilde c_{k\sigma}^\dagger 
\tilde c_{k\sigma}
\ege
\bge
H_B=U\sum_i d_i^\dagger d_i - 
\sum_{i\sigma} \lambda_{i \sigma} (s_{i\sigma}^\dagger s_{i\sigma} + d_i^\dagger d_i) +\sum_i \lambda_{i}^\prime
(1 - e_i^\dagger e_i - \sum_{\sigma}s_{i\sigma}^\dagger s_{i\sigma}
-d_i^\dagger d_i)
\ege
\bge
H_p=-\sum_{kk^\prime}V_{kk^\prime} \tilde q^2
\tilde c_{k\uparrow}^\dagger \tilde c_{-k\downarrow}^\dagger \tilde c_{-k^\prime \downarrow}
\tilde c_{k^\prime \uparrow}
\label{hp}
\ege
where, $\epsilon_k= -2t\left [ (cos k_x a + \gamma_1 cos k_y a) +\gamma_2 cos k_x a
cosk_y a \right ]$, where $\gamma_2 = \frac{2t^\prime}{t}$, t and $t^\prime$
represent nearest neighbour and next nearest neighbour hopping respectively,
$\gamma_1 = 1$ for square lattice and in presence of orthorhombic distortion 
$\gamma_1 < 1$. 

\par Since in the saddle point approximation the bosonic part $H_B$ is constant,
so the effective Hamiltonian is given by
\bge
H=H_0+H_{p}
\label{ham}
\ege
\par It can be seen from eqn. \ (\ref{hp})  that the pairing vertex between the correlated
electrons mediated by some bosonic exchange is renormalized to 
$\tilde q^2 V_{kk^\prime}$. 
Since $\tilde q^2$ deviates from unity ($\tilde q^2 <1$ for
$u \neq 0$), the SC-pairing amplitude will always be suppressed in presence
of electron correlation, however, its details will depend on the value of 
$u, \d$ and the nature of $V_{k,k^\prime}$, is the main point of investigation
in this work.
\par The SC order parameter in a correlated system may be
defined as
\bge
\Delta(k)=\sum_{k^\prime}\tilde q^2 V_{kk^\prime}
<\tilde c_{k^\prime \uparrow}^\dagger \tilde c_{-k^\prime\downarrow}^\dagger>
\label{mf}
\ege
The Hamiltonian \ (\ref{ham}) is treated within the mean field theory 
\ (\ref{mf}) in order to yield the SC gap equation within a weak coupling
theory as,
\bge
\Delta(k) = \sum_{k^\prime} \tilde q^2 V_{kk^\prime} \frac{\Delta(k^\prime)}
{2E_{k^\prime}} \tanh (\frac{E_{k^\prime}}{2T})
\label{gpeqn}
\ege
where the quasiparticle energy is given by
\bge
E_k^2=(\tilde q \epsilon_k - {\mu})^2+\mid \Delta(k) \mid^2
\ege
In the next section we will discuss about the different pairing
symmetry and its effect in a correlated system.

\section { Effect of electron correlation on order parameter symmetry }
Based on the nature of pairing potential the symmetry of 
SC gap could be different. The pairing potential is usually assumed
to have a separable form for simplicity i.e.,
\bge
V_{kk^\prime} = V \eta_k \eta_{k^\prime}
\ege
Depending on the nature of the $k$-dependence of $\eta_k$ one gets 
different symmetries such as
(i) $\eta_k$ = constant, corresponds to an isotropic $s$-wave 
(conventional BCS) 
symmetry, (ii) $\eta_k = f(k)$, refers to an anisotropic $s$-wave symmetry,
if $f(k)$ is a smooth function in the first Brillouin zone and is positive
definite (i.e., nodeless), (iii) $\eta_k = cos k_x a + cos k_y a$, corresponds
to an extended $s$-wave pairing symmetry and (iv) $\eta_k=cos k_xa - cos k_y a$
corresponds to $d_{x^2-y^2}$ pairing symmetry. In the following
we shall discuss only about the isotropic $s$-wave, the $d$-wave and a
mixed symmetry state which will be defined later on. 
\subsection{Pure $s$-wave}
The form of the pairing potential for isotropic $s$-wave symmetry is given by
\bge
V_{kk^\prime} = v_s = constant
\ege
and the corresponding SC gap function $\Delta_k \equiv \Delta_s$.
Hence the gap equation for a pure s-wave
\bge
\Delta_s = \tilde q^2 v_s \sum_{k^\prime} \frac{\Delta_s}{2E_{k^\prime}}
\tanh (\frac{\b E_{k^\prime}}{2})
\label{seqn}
\ege
The gap equation \ (\ref{seqn}) can be solved analytically following the
standard procedure of the BCS gap equation. However, we solve the gap equation
numerically as the same will not be possible to solve analytically in case
of $d$-wave as well as for the complex order parameter (discussed later on). 
All the parameters are expressed in units of `$t$'. However, at the end
temperature has been expressed in Kelvin assuming $t = 0.3$ eV (realistic
for copper oxide system), in order to examine, the real transition temperatures. For a set of parameter 
values $\omega_c=0.6$ (cut-off value), $v_s=1.0$ 
we present the numerical results 
in Fig. 1 and Fig. 2. 
Fig. 1 describes the temperature variation of the pure 
$s$-wave gap for a perfect
square lattice whereas the Fig. 2 represents that for a distorted lattice
($t_x \neq t_y$ , i.e., $\gamma_1 = 0.95$) with next nearest neighbouring
hopping $t^\prime=0.02t$. 
It is evident from both the figures 1 and 2 that the magnitude
of $\Delta_s$ as well as the corresponding $T_c$ decreases sharply with $u$.
Furthermore, for the same set of parameter not only that the magnitude of
$\Delta_s$ and $T_c$ is less in the distorted case (cf. Fig. 2) than that
for the perfect square lattice but also the fall in $T_c$ with increasing
$u$ is more in comparison to the former case. So, the suppressive effect 
of electronic 
correlation on $\Delta_s$ and $T_c$ is comparatively more in the orthorhombic 
phase (this may be clearer from Fig. 5) as will be discussed
in the later subsections.

\subsection{Pure $d$-wave}
In case of the pure $d_{x^2-y^2}$ symmetry the pairing potential is given by
\bge
V_{kk^\prime} = v_d
(cosk_x - \beta^\prime cos k_y)
(cosk_{x}^\prime - \beta^\prime cos k_{y}^\prime)
\ege
and the corresponding gap function is assumed as,
\bge
\Delta_{d}(k) = \Delta_{d}^{0} (cos k_x - \beta^\prime cos k_y)
\ege
where $v_d$ and $\Delta_{d}^{0}$ represents the strength of the $d$-wave pairing
potential and the amplitude of the gap function respectively. 
Where $\beta^\prime = 1$ or $< 1$ corresponding to a square lattice or a
orthorhombic lattice.  
So the resulting gap equation is written as
\bge
\Delta_d^{0}= \tilde q^2 v_d \sum_{k^\prime}\frac {\Delta_{d}^{0}(cos k_x^\prime 
-\beta^\prime cos k_y^\prime)^2 }{2E_{k^\prime}}
\tanh (\frac {\beta E_{k^\prime}}{2})
\label{deqn}
\ege
In Fig. 3 and Fig. 4 we show the thermal variation of the d-wave gap
for the set of parameter values $\omega_c=0.6$, $v_d=0.5$ for square 
($\gamma_1=\beta^\prime=1$) and
orthorhombic ($\gamma_1=\beta^\prime=0.95$) lattices
 with next nearest neighbour hopping
$t^\prime=0.02t$ respectively. The qualitative trend of behaviors 
for the thermal
variation of the gap parameter remains almost unaltered i.e, the magnitude
of the gap as well as the $T_c$ reduces with $u$. 
In case of orthorhombic lattice
the SC gap and $T_c$ is smaller than that of the undistorted one. 

Fig. 5 shows the variation of $T_c^0-T_c$ vs $u$ for $s$ and $d$ wave
symmetry in distorted and undistorted lattice. 
$T_c^0$ is the SC $T_c$ for $u=0$.
It is evident from Fig. 5 that for distorted lattice the effect of $u$ on 
$T_c$ is more than the undistorted one for both $s$ and $d$ wave symmetry.
It may be compared from figures 1 and 3 that for an increase in $u$ from 0
to 0.4 there is a reduction in $T_c$ about $50\%$ and $44\%$ for the $s$ and
the $d$ wave gaps respectively. The same when compared between the figures
2 and 4 the reduction in $T_c$ is about $58\%$ and $54 \%$ 
for the $s$ and $d$-wave respectively. Therefore, in general the coulomb
correlation 
affects the $s$-wave symmetry largely in comparison to the $d$-wave gap.
\subsection{Mixed $s+id$ order parameter symmetry}
In the mixed symmetry state there is a mixture of both $s$ and $d$ wave order
parameter and a phase difference of $e^{i\frac{\pi}{2}}$ is considered \cite{4.16}.
The generalized form of the pairing interaction causing the superconductivity is given by
\bge
V_{kk^\prime} = v_s + v_d(cos k_x - \beta^\prime cos k_y)(cos k_x^\prime-\beta^\prime cos k_y^\prime)
\label{mixv}
\ege
where $v_s$ and $v_d$ corresponds to the strengths of $s$ and $d$ wave 
channel interactions. And the corresponding $s+id$ 
wave order parameter is taken as,
\bge
\Delta(k) = \Delta_s + i \Delta_{d}^{0}(cosk_x - \beta^\prime cos k_y)
\label{mixop}
\ege
Substituting the equation \ (\ref{mixv})  and \ (\ref{mixop})  into equation
\ (\ref{gpeqn})  
and separating the real and imaginary parts of the equation we obtain 
gap equations for coupled $s$ and $d$ wave
components given by
\bge
\Delta_s = \tilde q^2 v_s \sum_{k^\prime} \frac {\Delta_s}{2E_{k^{\prime}}^{mix}}
\tanh (\frac {\beta E_{k^\prime}^{mix}}{2})
\label{mixs}
\ege
and 
\bge
\Delta_d^{0} = \tilde q^2 v_d \sum_{k^\prime}\frac {\Delta_d^{0}(cos k_x^\prime 
-\beta^\prime cos k_y^\prime)^2 }{2E_{k^\prime}^{mix}}
\tanh (\frac{\beta E_{k^\prime}^{mix}}{2})
\label{mixd}
\ege
with 
\bge
{E_{k}^{mix}}^2=(\tilde q \epsilon_k - {\mu})^2 + \Delta_s^2 + {\Delta_d^{0}}^2
(cos k_x - \beta^\prime cos k_y)^2
\ege
Self-consistent solution of equation \ (\ref{mixs})  and \ (\ref{mixd})  gives $\Delta_s$ and
$\Delta_d^{0}$ in a mixed $(s+id)$ wave state and for either $v_d=0$ or $v_s=0$ the coupled
gap equations reduce to pure $s$-wave or pure $d$-wave (as discussed in the
earlier subsections) respectively. To note, the difference between 
\ (\ref{seqn}) and \ (\ref{mixs}) as well as that between \ (\ref{deqn}) and
\ (\ref{mixd}) lies mainly with the difference between $E_k$ and $E_k^{mix}$.
Here we
performed 
the self-consistent solution of \ (\ref{mixs}) and \ (\ref{mixd}) 
with a cut off $\omega_c=0.6$ (as in the earlier cases of a pure $s$ or a $d$ wave scenario). 
\par In case of mixed $s+id$ symmetry with $v_d$ and $v_s$ both being 
finite, Fig.6
shows the phase diagram where the magnitudes of the order parameters with
different symmetry at a temperature of 10K are plotted against the ratio
of the strength of interactions 
($\frac{v_d}{v_s}$) for different $u$ for the square lattice. This phase
diagram has three regions, the region where only $s$-wave solution 
exists i.e, for $\frac{v_d}{v_s} <0.45$, only $d$-wave solution exists i.e, for 
$\frac{v_d}{v_s}>0.39$
and both the $s$ and $d$ wave solution co-exist for 
$0.39<\frac{v_d}{v_s}<0.45$.
For finite electron correlation $u=0.4$ the region with mixed symmetry shrinks
to
within $0.39<\frac{v_d}{v_s}<0.42$ with larger decrease in the magnitude
of the $s$ wave gap compared to that for the $d$-wave component.
To note, with the electron correlation increased (to $u=0.4$) the region
(in terms of $\frac{v_d}{v_s}$) where the $s$ -wave solution exists reduces
(to $\frac{v_d}{v_s} <0.42$), but
 that for the $d$-wave remains unchanged although
the magnitude is suppressed.
It is also important to note that the magnitudes of the $s$ or $d$- wave component gaps with correlation
depends crucially on the relative strength of interaction $\frac{v_d}{v_s}$.
 
\par Furthermore for $\frac{v_d}{v_s} \sim 0.39$ the $s$-wave
gap decreases with $u$ while $d$-wave gap increases slightly from its value 
compared to that at the uncorrelated
case. This is because the rate of decrease of $s$-wave gap is more rapid
than the $d$ wave gap with 
correlation. For a clearer observation, the variation in the temperature 
dependence of the $s$-wave and $d$-wave component of the
gap parameters in the mixed state (for $\frac{v_d}{v_s} = 0.4$) 
with electronic 
correlation is shown in Fig. 7. The graph with diamond shows pure s-wave gap
for the uncorrelated case ($v_s=1$, $v_d=0$). In the mixed state $\frac
{v_d}{v_s}=0.4$, (and in the absence of correlation) 
the s-wave gap has value slightly less than that of the pure 
case (the dashed line). With 
increasing correlation, $u=0.4$, while the magnitude of s-wave gap is suppressed
strongly (dotted line),
 that of the d-wave gap acquires higher value than the $u=0$ 
case (the solid line).
Nevertheless, the d-wave $T_c$ ($T_{d}^c$) is higher
for the uncorrelated case than that with correlation. Hence, in the lower 
temperature regions there is a competing effect between the $s$ and $d$-wave 
gap parameters
in the mixed symmetry region; with the appearance of the $s$-wave 
order parameter the $d$-wave gap is suppressed 
for the $u=0.4$ case (the dash dotted line). 
The $d$-wave gap attains its highest value where the $s$-wave component
gap amplitude vanishes. In this figure, the curve with $++$ sign represents the
pure $d$-wave gap with all the parameters being same ($v_d = 0.4$) ; demonstrating the
arresting of the growth of the $d$-wave component with the opening up of the
$s$-component gap in the mixed phase.

\par Figures 8 and 9 show the variation of $s$-wave and $d$-wave
gap parameters for different $u$ as the ratio of the strengths of interactions
increases  close to and 
beyond the values corresponding to the co-existence regions (for $u =0.4$)
 i.e., 
at $\frac{v_d}{v_s}=0.42$ and $\frac{v_d}{v_s}
=0.45$ respectively. The gap magnitude for both the order parameter symmetry
decreases with electron correlation, however, the rate of decrease is different
for $s$ and $d$ in the mixed state which depends on $\frac{v_d}{v_s}$ ratio
as well as the electron correlation. Here for the uncorrelated case it
is clearly visible that the growth of $d$-wave component of the gaps is
arrested with the onset of the $s$-wave component gap. 
In fact a detailed investigations of the Figures 7 to
9 will  demonstrate that in the mixed phase within the coexistence regime
i.e ($0.39<\frac{v_d}{v_s}< 0.42$),
the suppression in different components of the gap as well as the 
corresponding $T_c $s are very sensitive to the relative pairing strengths
of $s$ and $d$ channels. 
\par So far all the calculations for the mixed phase, were carried out for
the square lattice. However, it was seen in case of the pure $s$ and $d$
phases that the introduction of orthorhombicity causes reduction in both,
the magnitude of the order parameters as well as the corresponding $T_c$s.
So the effect of orthorhombicity on the mixed phase is now investigated.
The phase diagram for different order parameter symmetry 
as a function of the $\frac{v_d}{v_s}$ ratio is depicted in the Fig. 10 for the orthorhombic
distortion and finite next nearest neighbouring interaction. The $s$-wave
and $d$-wave gap has lesser magnitude in the distorted lattice than that 
for the perfect square
lattice in mixed $s+id$ state and the region of $s$ and $d$ co-existence 
is shifted to 
smaller $\frac{v_d}{v_s}$ value and is shrunk to smaller
region of area compared to the undistorted lattice. 
Similar to the phase diagram in Fig. 6, the value of $\frac{v_d}{v_s}$
for the occurence of $d$-wave component does not change with $u$ whereas that
for the $s$ component changes from $\frac{v_d}{v_s}<0.48$ (for $u=0)$ to
$\frac{v_d}{v_s} < 0.46$ (for $u = 0.4$). With 
a higher value of Coulomb correlation ($u = 0.4$) the $s-d$ mixing region
reduces further along with a large suppression in the magnitude of $s$-wave and 
$d$-wave gap parameters.
It is noticable that the $s$-wave component suffers larger
suppression in magnitude of the gap than that for the $d$-wave. 
The on-site Coulomb
correlation strength $u$ as well as the isotropic pairing strength
$v_s$ both being isotropic in nature, the $s$-wave gap magnitude is largely
affected by Coulomb correlation, whereas due to extended nature of the 
$d$-wave component it is less affected by correlation.
However, we note that in the mixed phase, depending on the 
value of $\frac{v_d}{v_s}$ the effect of electron correlation
on transition temperatures 
$T_c^d$ or $T_c^s$ is very different.
This is due to the fact that in the mixed phase
none of the components ($s$ or $d$) follow BCS like temperature variation.
This has sometime led to increase in the gap magnitude with $u$
but reduction in the $T_c$ ($\equiv T_c^d$) (cf. Fig. 7). 
\par Fig. 11 shows the variation of $s$ and $d$ wave component 
with electron correlations for
$\frac{v_d}{v_s}=0.45$. Similar to the square lattice the $s$ and $d$ wave gap
parameters in the pure state take higher values than that 
in the mixed state for 
same parameter values ${v_d}$, ${v_s}$ and $u$. 
This figure has to be contrasted
with Fig. 9. 
With the introduction of orthorhombicity the nature of the temperature
variations of the different gap components changes abruptly. It is 
worth noting that in the pure $s$ or pure $d$ wave cases (discussed earlier)
the orthorhombicity suppresses both the magnitude of 
the order parameters and the corresponding
$T_c$s (cf. Fig. 1 to Fig. 4). However in the mixed state,
because of the interplay between the two components of the order
parameters, the orthorhombicity together with next nearest neighbour 
hopping enhances the relative magnitude of the $s$-wave component whereas
reduces that of the $d$-wave when compared with respect to the perfect
square lattice for same value of $\frac{v_d}{v_s}$
(cf. figures 9 and 11). 
Considering
the case of $u=0.4$ from Fig. 9 and Fig. 11 it would be 
self-evident that the $T_c^d$ in the orthorhombic phase reduces very
largely from its value for the square lattice, whereas $T_c^s$ in the
orthorhombic phase increases strongly. On the other hand  unlike the
undistorted lattice, with the mixed symmetry (cf. Fig. 6), where the
$d$-wave component can have higher magnitude for higher $u$ close to 
$\frac{v_d}{v_s}$=0.4 ; in  presence of orthorhombic distortion the $d$-wave
order parameter never acquire higher magnitude for higher $u$,  
for any value of $\frac{v_d}{v_s}$, as is clear from Fig. 10
as well as Fig. 11.
\section { Temperature dependent gap anisotropy in a correlated system}
There are three broad catagories of experiments, which had been used to probe
the symmetry of the superconducting gap function 
\cite{4.8,4.9,4.10,4.19,4.17,4.18} namely 
(i) magnetization and transport, (ii) spectroscopies and excitations and
(iii) Josephson measurements. The most direct and fundamental probe of the
gap magnitude in superconductors is angle resolved photoemission spectroscopy
(ARPES), which indicates a highly anisotropic gap in high $T_c$ compounds.
\par The main features observed from ARPES by different groups of 
experimentalists are summarized below
 : Shen et al, \cite{4.18} observed that the
SC-gap has (a) maximum magnitude along $\Gamma - M$ direction (Cu-O bond
direction in real space) ; (b) minimum magnitude (negligible spectral weight)
along $\Gamma - X(Y)$ direction (diagonal to the Cu-O bond) and there is a
monotonic increase in the magnitude of the gap from its smallest value along
the $\Gamma - X$ direction, thereby confirming true anisotropy. The authors
of \cite{4.18} argue that these results are consistent with $d_{x^2-y^2}$ 
symmetry of the gap. However, the data on the momentum - resolved temperature
dependence of the SC gap of Bi-2212 by J. Ma et al \cite{4.11} is not only
in contradiction with Shen et al \cite{4.18} but also it shows some exciting
features. For measurements of the temperature dependence of the gap, the angle
resolved photoemission data have been taken for temperatures ranging from 36 to 95 K
along the $\Gamma - M$ as well as in the $\Gamma - X$ (40 to 95 K) 
high symmetry directions. The photoemission SC condensate spectral area along
$\Gamma - M$ direction exists till the bulk transition temperature, $T_c=83$K.
The measured SC gap is almost independent of the temperature at lower 
temperatures and retains its maximum value even up to temperature 85$\%$ of its
$T_c$. But most interesting features to note is that, the photoemission
condensate spectral area is remarkably weaker at 40K in the $\Gamma - X$
direction compared to that in the $\Gamma - M$ direction, indicating a smaller
size of the gap. The size of the gap at 36 K in the $\Gamma - M$ direction
is 16 meV whereas that in the $\Gamma - X$ direction is 10 meV (at T=40 K).
The finite gap magnitude along $\Gamma - X$ direction below $T_c$ ruled out
the possibility of simple $d$-wave pairing \cite{4.19}. However, at 70 K
when the gap along the $\Gamma - M$ direction retains 90 to 100$\%$ of
its full value, becomes indistinguishable from zero along the $\Gamma - X$
line. This may be taken as a signature of a two component order parameter,
$d_{x^2-y^2}$ type close to $T_c$ and a mixture of the both $s$ and $d$ otherwise.
Moreover, the significant point is that, due to the unconventional temperature
dependence of the SC gap in different parts of the Brilloun zone, the
temperature dependent gap anisotropy i.e., the ratio 
$\frac{\Delta_{sc}(\Gamma - M)}{\Delta_{sc}(\Gamma - X)}$ is enhanced 
closed to $T_c$.
In the present section we show that the above peculiarity in the 
temperature dependence of the SC gap is a natural consequences of the mixed
$s+id$ symmetry and in a correlated system the gap anisotropy increases with
electron correlation. 
\par In the gap equations \ (\ref{mixs}) and \ (\ref{mixd})  
$\Delta_s$ and $\Delta_{d}^{0}$
represents the amplitudes of the $s$ and $d$ wave gaps respectively. Since $\Delta_s$
represents the isotropic $s$-wave gap so it retains its constant maximum value
for any $k_x$ and $k_y$ direction whereas magnitude of $\Delta_d$
varies with $k_x$ and $k_y$ direction due to the multiplicative factor
($cos k_x - cos k_y$). In the $\Gamma - \bar M (M)$ direction i.e., ($\pm \pi$, 0),
(0,$\pm \pi$), $(cos k_x - cos k_y)$ acquires maximum magnitude $\mp 2$,
while in the $\Gamma - X(Y)$ direction ($\pm \frac{\pi}{2}$, $\pm \frac{\pi}{2}$)
($cos k_x - cos k_y$) is zero identically. So it is expected that the $d$-wave
gap parameter will attain its maximum at $\Gamma - M (\bar M)$ points and minimum
at $\Gamma - X(Y)$ points, however $s$-wave gap is constant all 
over the Fermi surface. Considering the energy spectrum of the mixed
phase $E_{k}^{mix}$ we write $\Delta(\Gamma - M)$ and $\Delta (\Gamma - X)$ as
\bge
\Delta (\Gamma - M) = \sqrt{\Delta_s^2 + 4 \Delta_d^2}
\ege
and
\bge
\Delta (\Gamma - X) = \Delta_s
\ege
Hence the ratio which is a measure of gap anisotropy can be written as
\bge
\frac{\Delta(\Gamma - M)}{\Delta(\Gamma - X)} = \frac {\sqrt{\Delta_s^2
+ 4\Delta_d^2}}{\Delta_s}
\ege
Fig. 12 shows the variation of $\Delta (\Gamma - M)$ and $\Delta (\Gamma - X)$
for $u=0$ and $u=0.4$ in an orthorhombic lattice with $\frac{v_d}{v_s} = 0.46$.  The $\Delta (\Gamma - M)$ is larger and
 almost constant with temperature
till $\Delta (\Gamma - X)$ is zero, then $\Delta (\Gamma - M)$ falls to zero
at the $T_c$. The decrease of $\Delta (\Gamma - M)$ with $ u$ is much less
than that for the $\Delta (\Gamma - X)$. 
The inset figure of Fig. 12 depicts the temperature dependent gap
anisotropy in a correlated case. The ratio $\frac{\Delta(\Gamma - M)}{\Delta
(\Gamma - X)}$ is almost constant at lower temperatures and increases faster near
the temperature where $\Delta(\Gamma - X)= \Delta_s $ is going to be zero.
$\frac{\Delta(\Gamma - M)}{\Delta(\Gamma - X)}$ ratio is higher for higher
correlation. This feature is however not an unique consequence of $s+id$ 
pairing symmetry e.g,
similar results are also obtained in a modified version of spin bag \cite{4.20}
mechanism of superconductivity.
\section{Conclusion}
We have presented a detailed study of the effect of electron correlation
on pairing 
symmetry within a weak coupling theory. The method used to treat electron
correlation is the slave boson formalism of KR [18] which reproduces
the Brinkmann-Rice value of metal-insulator transition ($U_c$) correctly
in the paramagnetic state. The value of $u$  used in this work is always
less than 1 (i.e $U<U_c$) so that weak coupling mean field theory of
superconductivity can be
applied. All through out the present calculation we restricted to
half-filled situation. 
Within this study we found that
electron correlation has important effect on pairing symmetry. The detailed
nature of order parameter, its temperature variation as well as magnitude
depends very sensitively with electron correlation and the nature of
underlying lattice considered. Few interesting features of interplay
between
 the different components of the order parameter with electron correlation 
has been demonstrated (in the $s+id$ picture). Such study may have some bearings 
to the high temperature superconductors. However, we have made no effort
to understand the mechanism of superconductivity for cuprates in this work.
The electron correlation always affect the pure $s$ wave gap more than the
pure $d$ wave symmetry gap irrespective of the underlying lattice.
In the mixed $s+id$ phase, the suppression in different components of the gap 
as well as the corresponding $T_c$ are very sensitive to 
$\frac{v_d}{v_s}$ and the underlying lattice.
Assuming the $s+id$ nature of the order parameter, the measured temperature
dependent gap anisotropy from the ARPES of J. Ma et al, [11] can be understood
on which the role of electron correlation has been emphasised. However,
such a picture is not unique [24].

\eject

 {\bf Figure Captions} \\

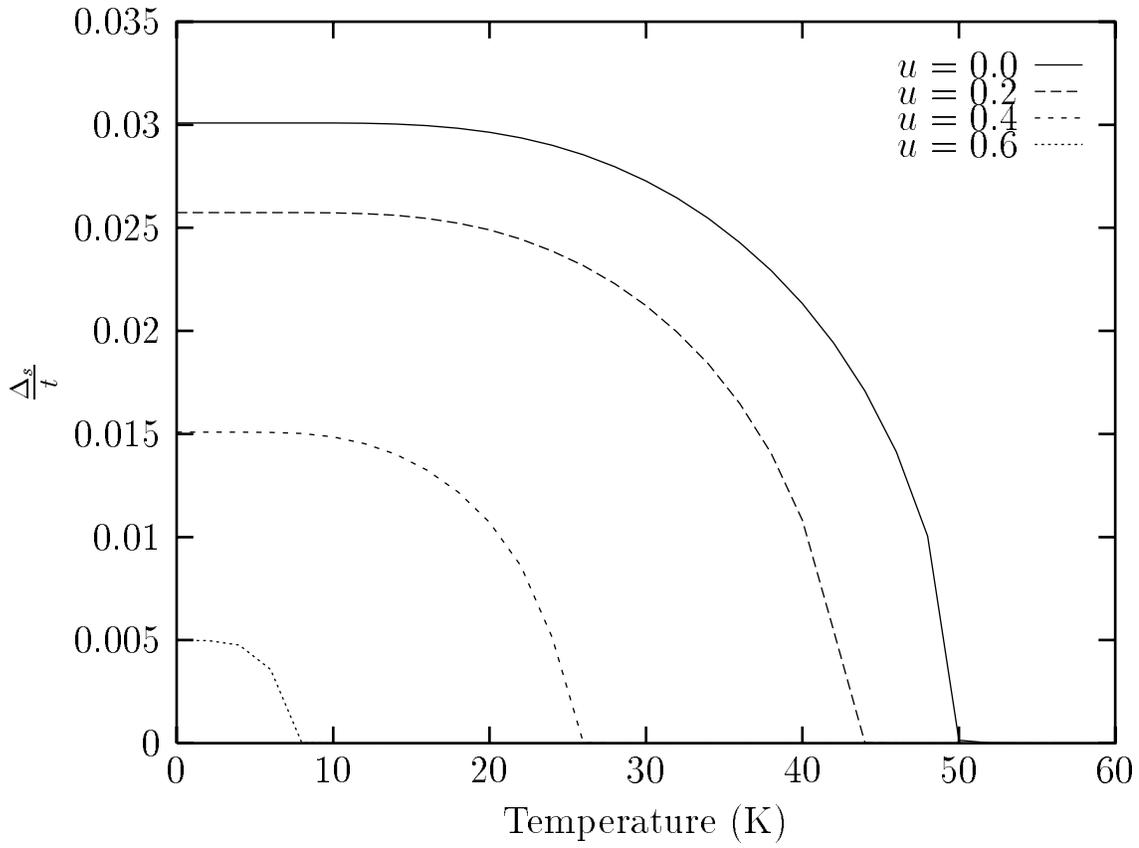
\noindent {\bf Fig.  1}  Thermal variation of the isotropic $s$-wave
gap ($v_s = 1$) for various values of the electron correlation 
($u = \frac{U}{U_c}$) in a
perfect square lattice (for other parameters, see text). \\

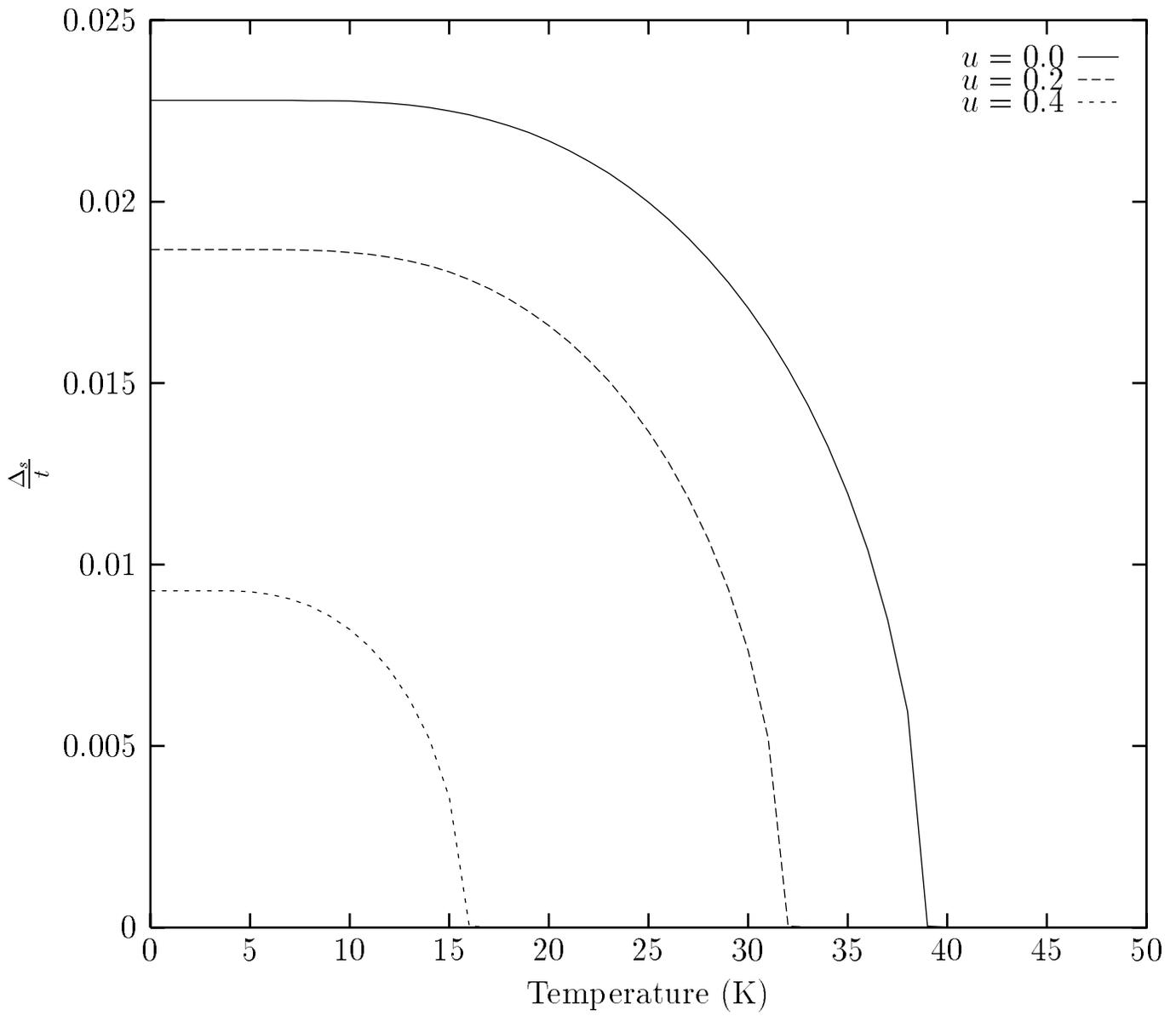
\noindent {\bf Fig.  2}  Thermal variation of the isotropic $s$-wave
gap ($v_s = 1$) in an orthorhombic lattice for various values of $u$
(all other parameters being same as that in Fig. 1). \\

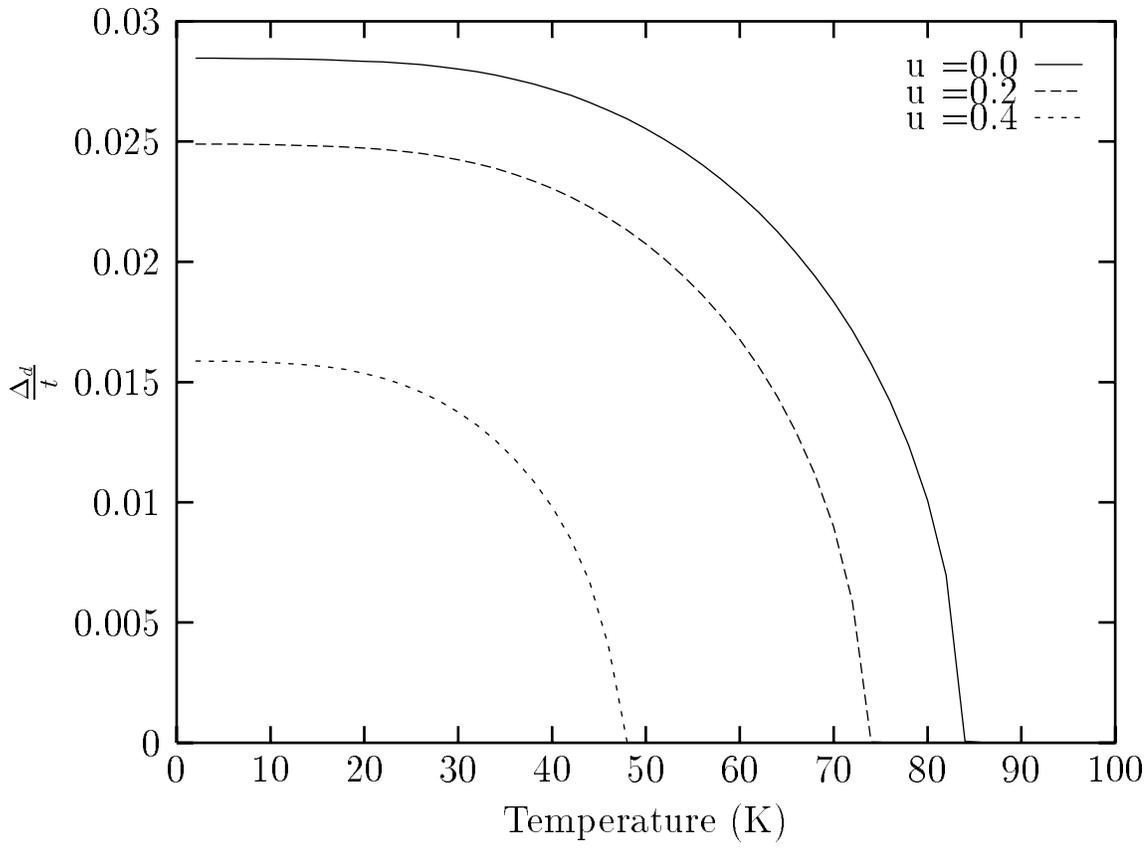
\noindent {\bf Fig.  3}  Thermal variation of the pure $d$-wave
gap ($v_d = 0.5$) for various values of the electron correlation
in a perfect square lattice (for other parameters see text). \\
 
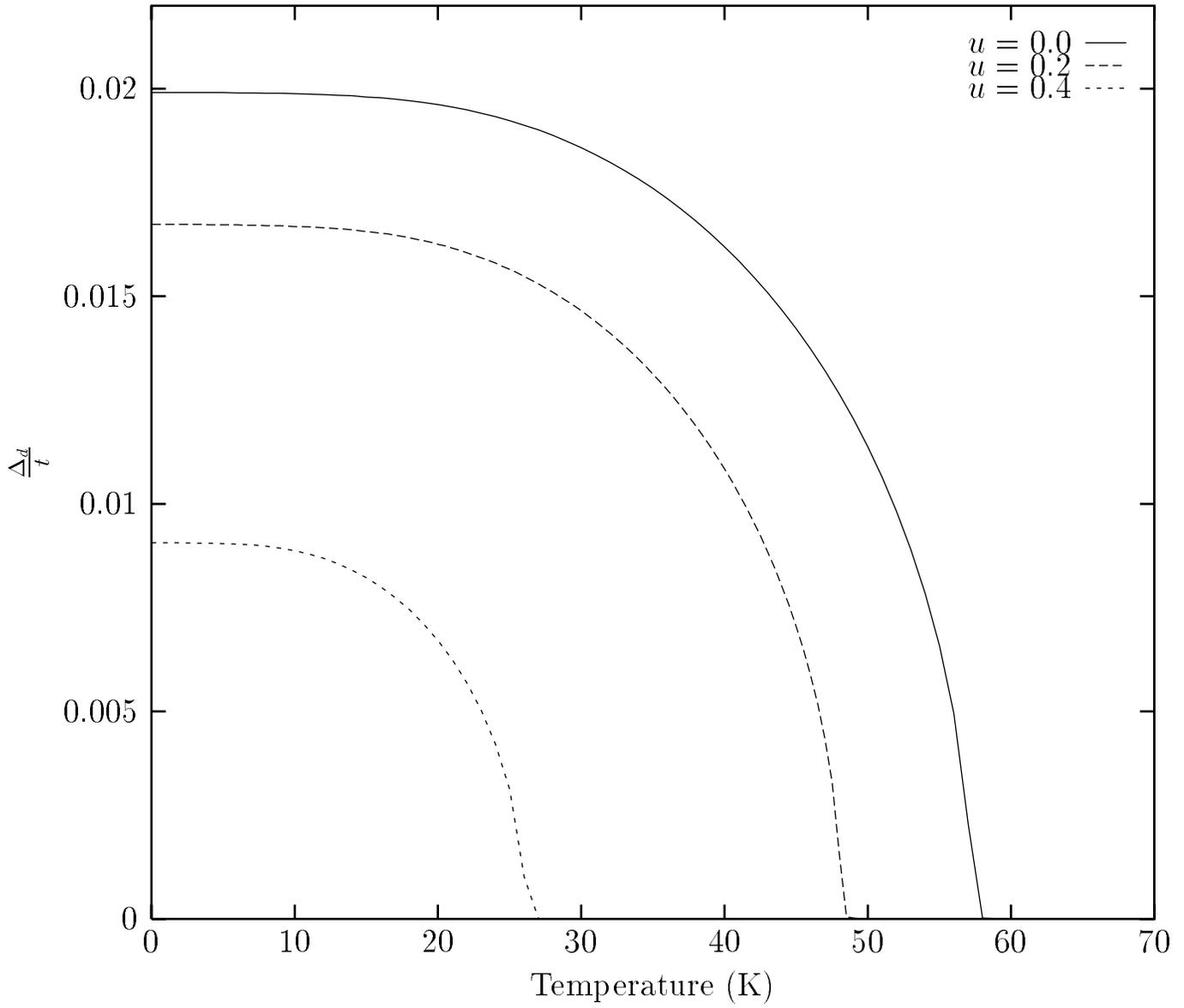
\noindent {\bf Fig.  4}  Thermal variation of the pure $d$-wave
gap ($v_d = 0.5$) for various values of electron correlation in
an orthorhombic lattice (with all the parameters being same as
Fig. 3). \\
 
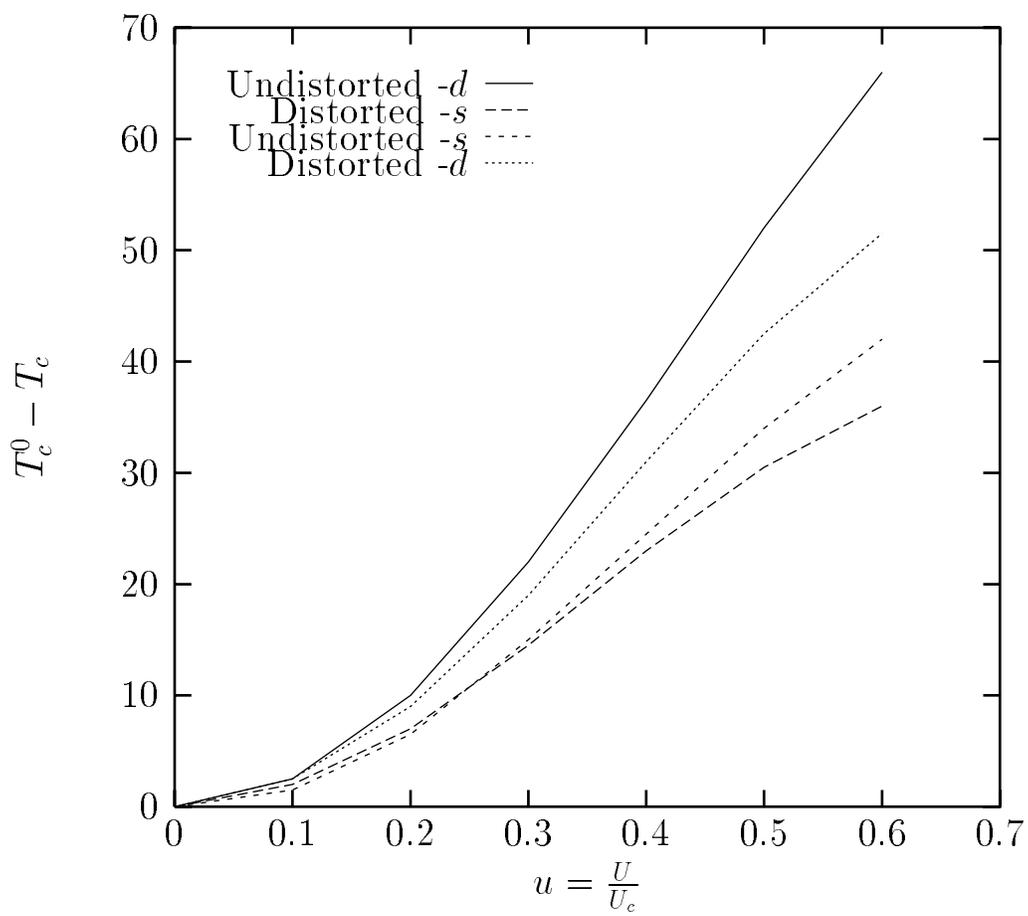
\noindent {\bf Fig. 5}  Change in  $T_c$ due to correlation
for  pure-$s$ and
pure-$d$ wave gaps (in square and orthorhombic lattices) where
$T_{c}^0 \equiv T_c ( u = 0)$. \\

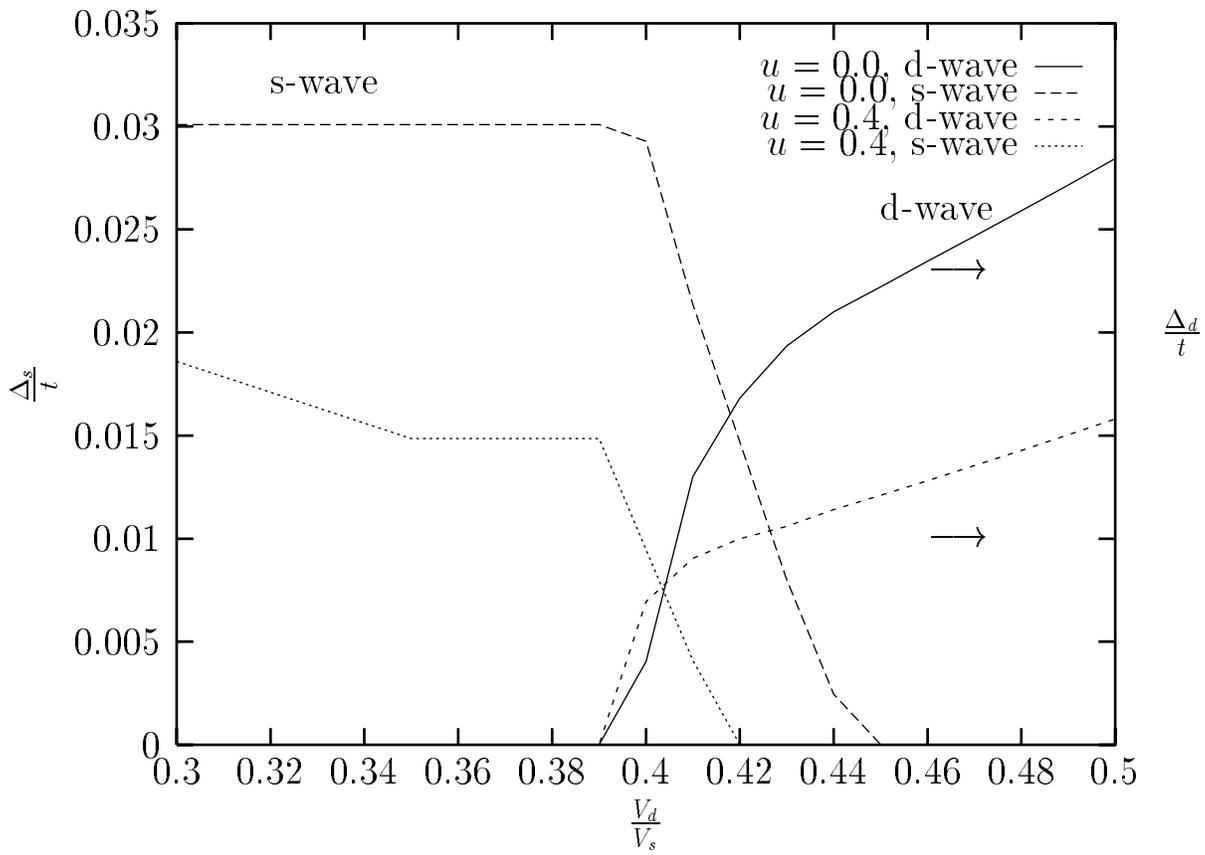
\noindent {\bf Fig. 6}  Phase diagram of a $s + id$ superconductor, 
where the amplitudes of the $s$-wave and $d$-wave
component of the  order parameter in a square lattice is plotted
 as a function of their relative strength $\frac{v_d}{v_s}$ for 
different values of electron correlation at T = 10 K. \\

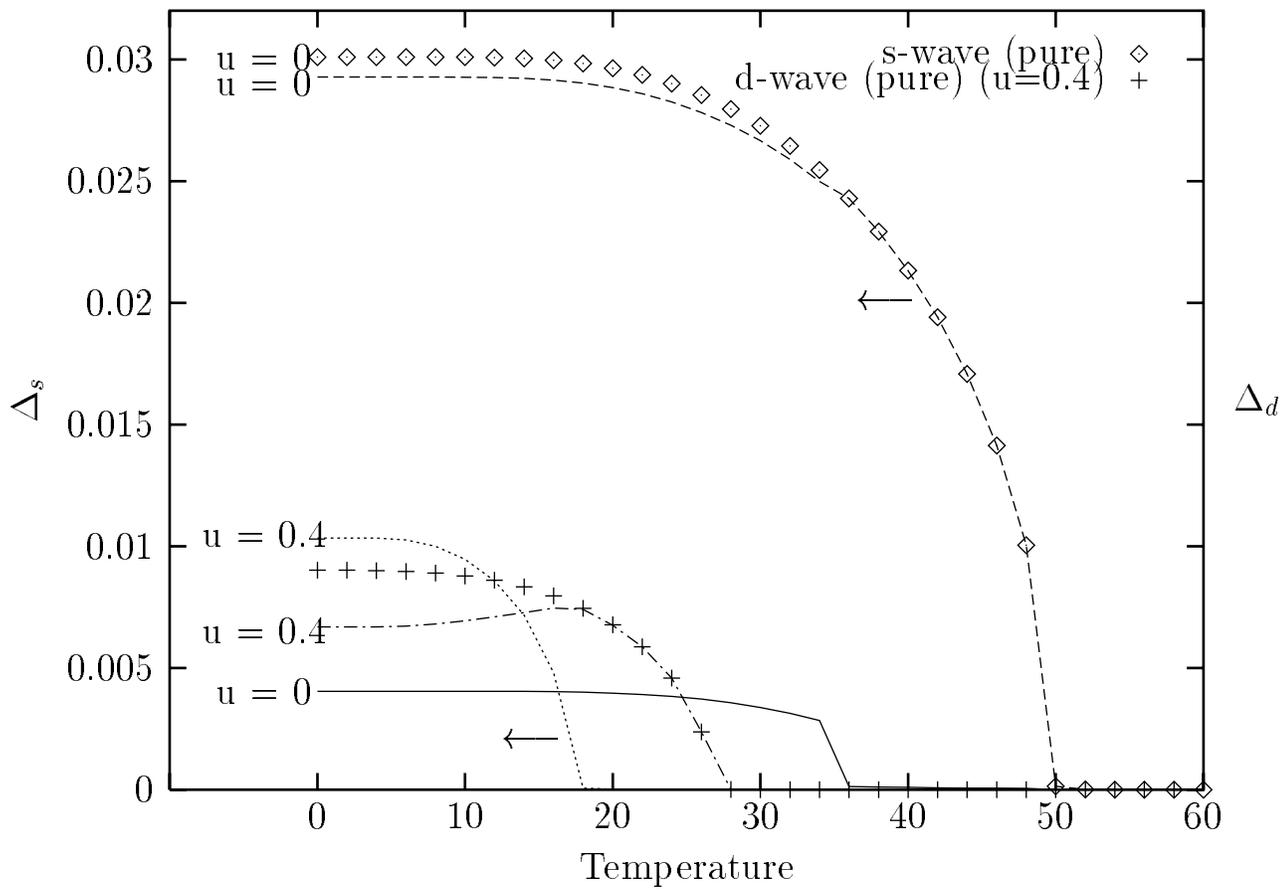
\noindent {\bf Fig. 7}  Thermal variation of the amplitudes
 $\Delta_s$ and $\Delta_d$ (in units of $t$) 
of the complex order parameter ($s +id$)
for different values of electron correlation at $\frac{v_d}{v_s}$ = 0.4 
in a perfect square lattice (all other parameters are kept fixed 
as earlier figures).\\
 
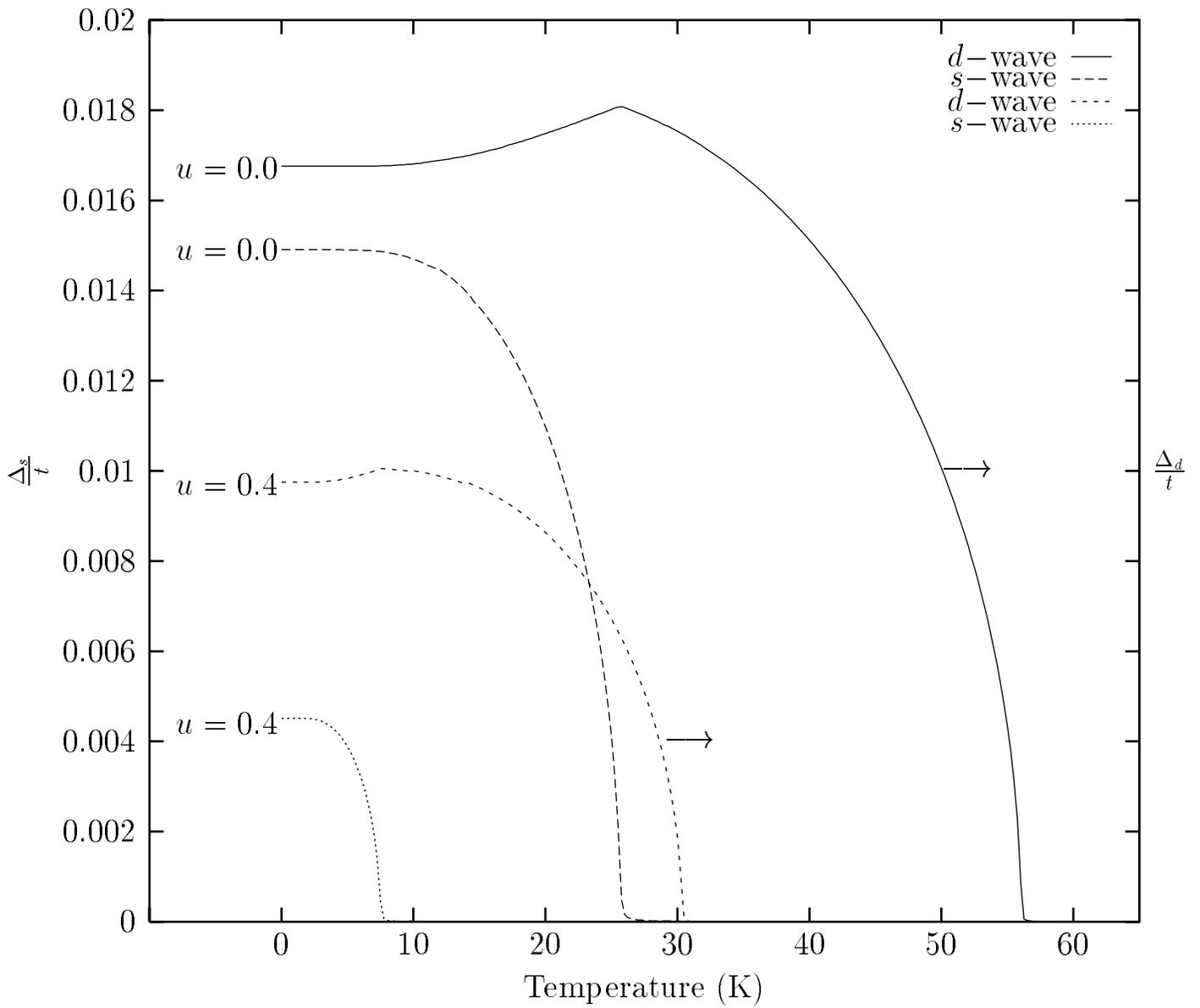
\noindent {\bf Fig. 8} Thermal variation of $\Delta_s$ and $\Delta_d$ 
 of the complex order parameter ($s +id$)
for different values of electron correlation at $\frac{v_d}{v_s}$ = 0.42
in a perfect square lattice (all other parameters are same as earlier
figures). \\

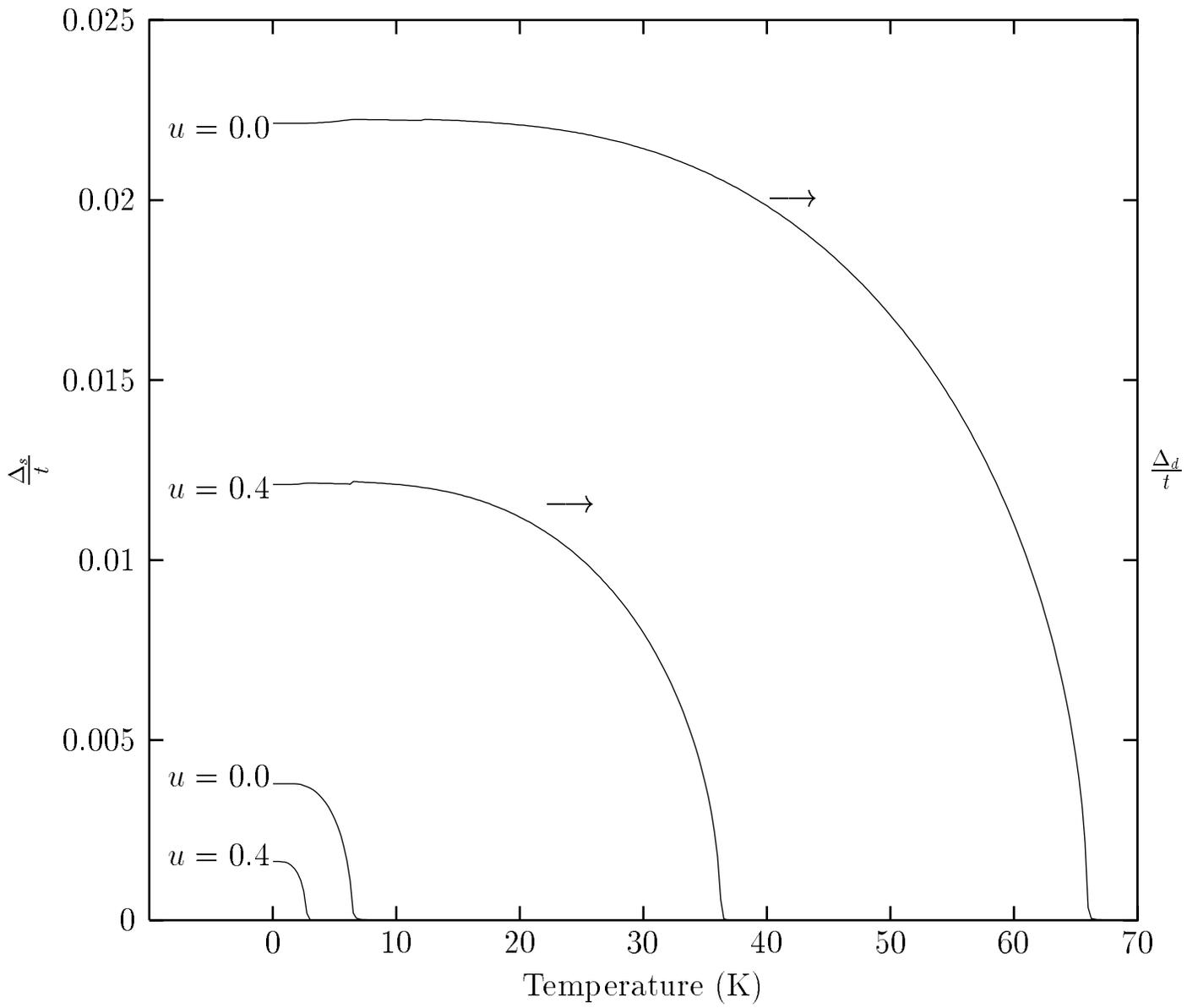
\noindent {\bf Fig. 9}  Thermal variation of $\Delta_s$ and $\Delta_d$ 
 of the complex order parameter ($s +id$)
for different values of electron correlation at $\frac{v_d}{v_s}$ = 0.45
in a perfect square lattice. \\

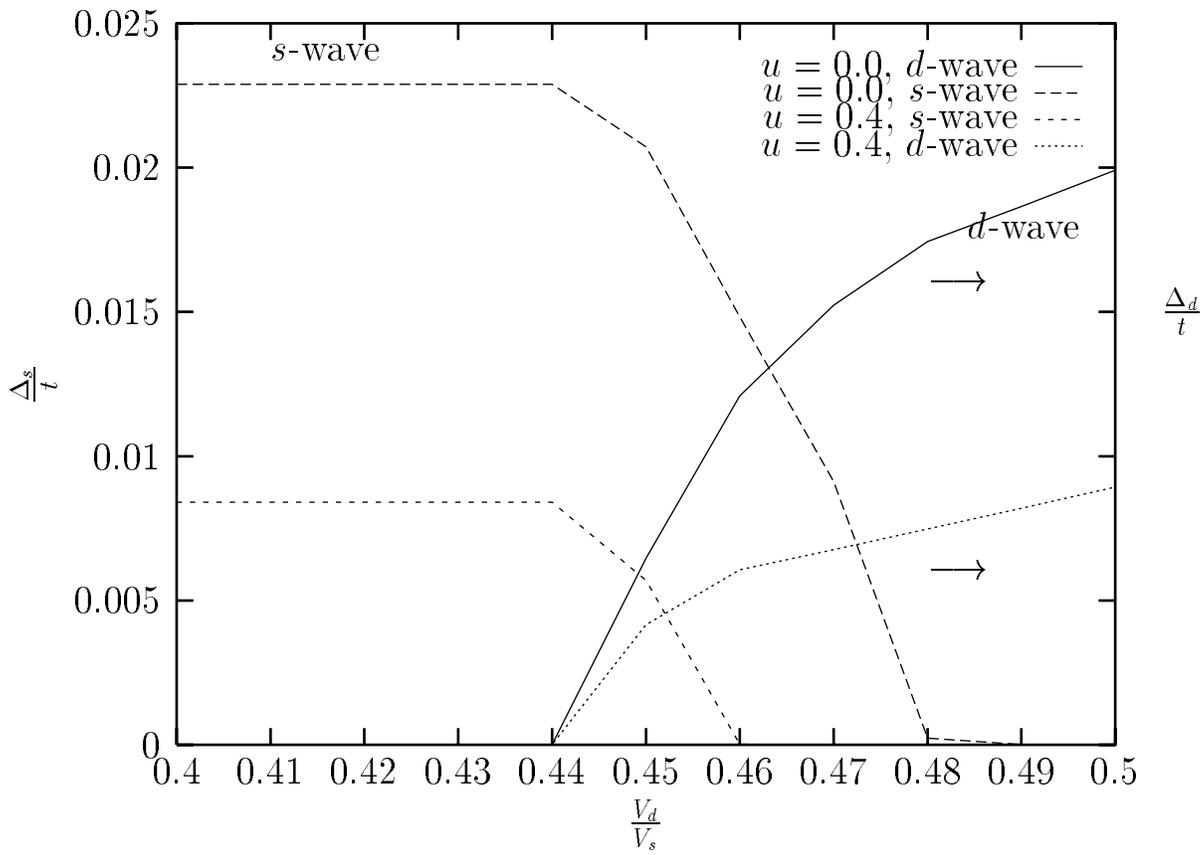
\noindent {\bf Fig. 10} Phase diagram of a $s +id$ superconductor where
the amplitudes  $s$-wave and $d$-wave
component of the  order parameter is drawn against their relative
pairing strength in an  orthorhombic lattice for different
electron correlations (T = 10K). \\
 
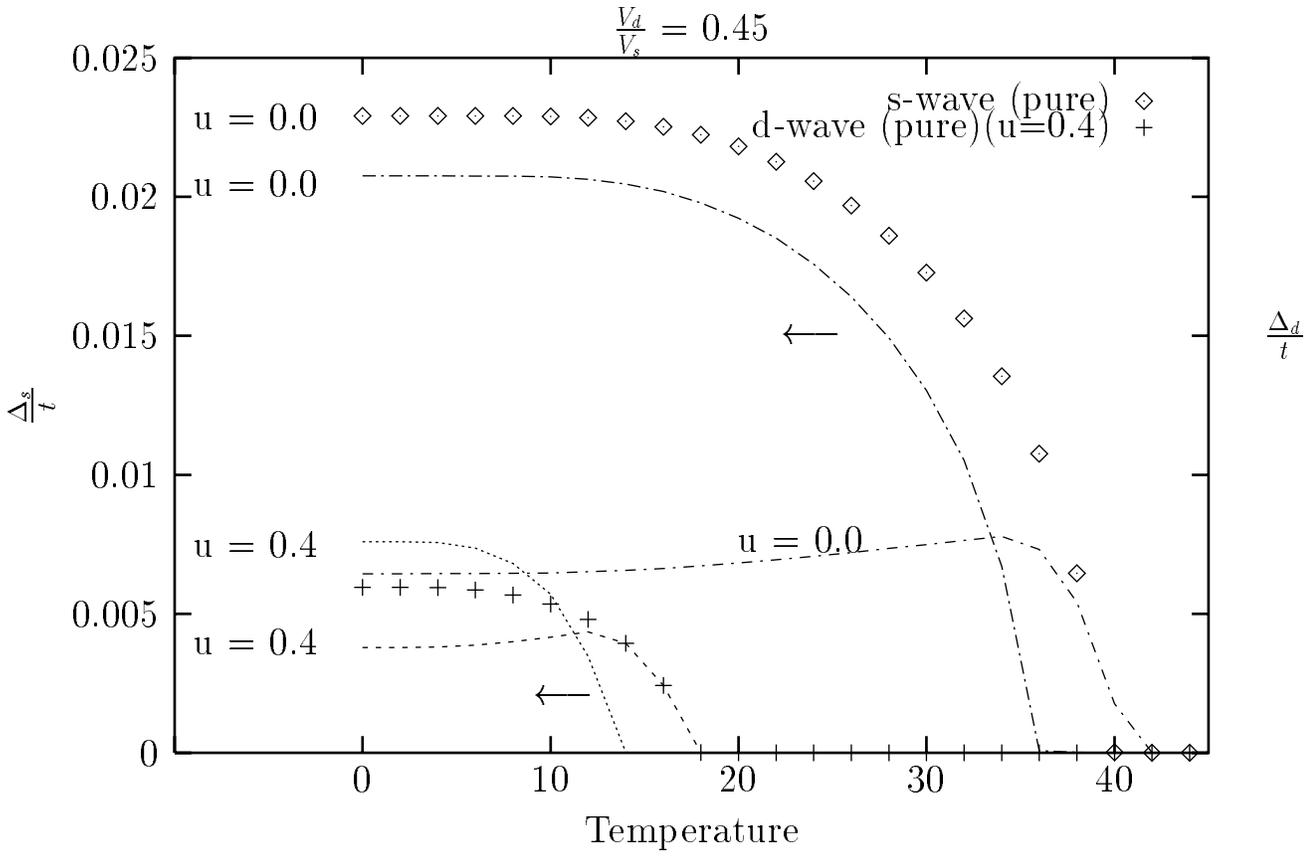
\noindent{\bf Fig. 11} Thermal variation of $\Delta_s$ and $\Delta_d$
of the complex order parameter ($s +id$)
for different values of electron correlation in an orthorhombic lattice at
$\frac{v_d}{v_s}$ = 0.45. \\

\noindent{\bf Fig. 12} Temperature variation of the superconducting
gap along the two high symmetry directions for $u$ = 0.0, 0.4
in the mixed $s+id$ phase. The inset figure depicts the ratio of the 
superconducting gap along the two high symmetry directions
$(\frac{\Delta_{sc}(\Gamma - M)}{\Delta_{sc}(\Gamma - X)})$ with
temperature for u = 0.0, 0.4. \\


\begin{thebibliography}{99}
\bibitem{4.1} C. E. Gough, M. S. Colcough, E. M. Forgan, R. G. Jordan, M. 
Keene, C. M. Muirhead, A. I. Rae, N. Thomas, J. S. Abell and S. Sutton,
Nature, {\bf 326}, 855(1987).
\bibitem{4.2} T. J. Witt, Phys. Rev. Lett. {\bf 61}, 1423(1988).
\bibitem{4.3} A. F. Andreev, Sov. Phys. JETP {\bf 19}, 1228(1964) ; 
G. ER. Blonder, M. Tinkham and T. M. Klapwijk, Phys. Rev. B {\bf 25},
4515(1982).
\bibitem{4.4} J. R. Kirtley, Int. J. Mod. Phys. {\bf 4}, 201(1990) ; T. Ekino
and J. Akimitsu in ``Frontiers in solid state sciences'', Vol. 1,
(Superconductivity), Eds : L. C. Gupta and M. S. Multani,(World Scientific
Publishers, Singapore, 1993), p.477
\bibitem{4.5} D. E. Maclaughlin in ``Solid State Physics", Vol. 31,
Eds : H. Ehrenreick, F. Seitz and D. Turnbull (Academic Press, New York, 1976),
p.2.
\bibitem{4.6} S. E. Barret, J. A. Martindale, D. J. Durand, C. P. Pennington,
C. P. Slichter, T. A. Friedmann, J. P. Rice and D. M. Ginsberg, Phys. Rev.
Lett. {\bf 66}, 108(1991).
\bibitem{4.7} J. A. Martindale, S. E. Barret, K. E. O'Hara, C. P. Slichter,
W. C. Lee and D. M. Ginsberg, Phys Rev. B {\bf 47}, 9155(1993).
\bibitem{4.8} W. N. Hardy et al, Phys. Rev. Lett. {\bf 70}, 3999(1994).
\bibitem{4.9} Dong Ho Wu et al, Phys. Rev. Lett. {\bf 70}, 85(1993).
\bibitem{4.10} A. G. Sun, D. A. Gajewski, M. B. Maple and R. C. Dynes,
Phys. Rev. Lett. {\bf 72}, 2267(1994).
\bibitem{4.11} J. Ma, C. Quittman, R. J. Kelley, H. Berger, G. Margaritando
and M. Onellion, Science {\bf 267}, 862(1995), H. Ding, J. C. Campuzano and 
G. Jennings, Phys. Rev. Lett. {\bf 74}, 2784(1995).
\bibitem{4.12} G. Kotliar, Phys. Rev. B {\bf 37}, 3664(1988).
\bibitem{4.13} A. E. Ruckenstein, P. Hirschfeld and J. Apel, Phys. Rev. B
{\bf 36}, 857(1987).
\bibitem{4.14} Q. P. Li, B. E. C. Koltenbah and R. Joynt, Phys. Rev. B
{\bf 48} 437(1993).
\bibitem{4.15} J. H. Xu et al, Phys. Rev. Lett. {\bf 73}, 2492(1994).
\bibitem{4.19} P. Monthoux and D. Pines, Phys. Rev. B {\bf 49} 4261(1994) ;
D. J. Scalapino, Phys. Rep. {\bf 250} 329(1994).
\bibitem{bcs} J. Bardeen, L. N. Cooper and J. R. Schrieffer, Phys. Rev.
{\bf 108} 1175(1957).
\bibitem{kot} G. Kotliar and A. E. Ruckenstein, Phys. Rev. Lett. {\bf 57},
1362(1986); M. Lavagna, Phys. Rev. B {\bf 41}, 142 (1990).
\bibitem{a} Manidipa Mitra, thesis submitted to the Utkal University, India
(1997).
\bibitem{b} M. Mitra and S. N. Behera, Cond. Matt. Mat. Commn. {\bf 1}, 247
(1994) ; P. Entel et al., Int. Jr. Mod. Phys. {\bf 5}, 271 (1991).
\bibitem{4.16} M. Liu and D. Y. Xing and Z. D. Wang, Phys. Rev. B. {\bf 55}
3181(1997).
\bibitem{4.17} S. N. Mao et al, Appl. Phys. Lett. {\bf 64}, 375(1994) ;
A. G. Sun et al Phys. Rev. Lett. {\bf 72} 2632(1994).
\bibitem{4.18} Z. X. Shen and D. S. Dessau, Phys. Rep. {\bf 253},
1(1995) ; Z. X. Shen et al, Phys. Rev. Lett. {\bf 70} 1553(1993).
\bibitem{4.20} Haranath Ghosh, Physica C {\bf 269} 55(1996).
\bibitem{4.21} C. O'Donovan and J. P. Carbotte, Physica C {\bf 252},
87(1995).
\end{thebibliography}
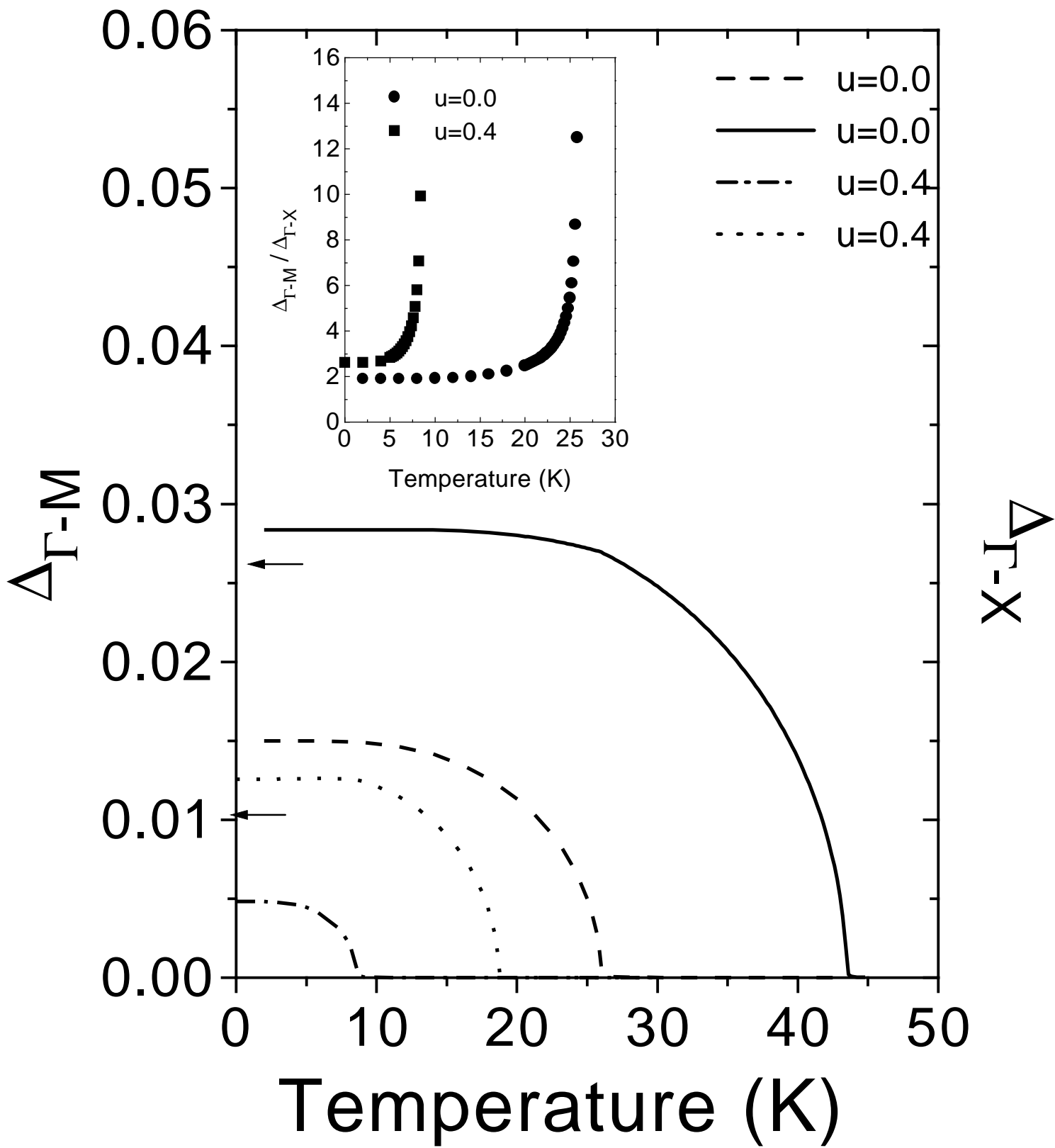
\end{document}